\documentclass{article}

\usepackage[main,final]{neurips_2025}
\usepackage{graphicx}

\workshoptitle{None}

\usepackage[utf8]{inputenc} 
\usepackage[T1]{fontenc}    
\usepackage{hyperref}       
\usepackage{url}            
\usepackage{booktabs}       
\usepackage{amsfonts}       
\usepackage{nicefrac}       
\usepackage{microtype}      
\usepackage{xcolor}         

\title{Diffusion Modeling of the Three-Dimensional Magnetic Field in the Sun's Corona}

\author{%
  Daniel E. da Silva \\ 
  Heliophysics Sciences Division\\
  NASA Goddard Spaceflight Center\\
  Greenbelt, MD 20770 \\
  \texttt{daniel.e.dasilva@nasa.gov} \\
    \AND
   Michael Kirk \thanks{Co-authors contributed equally.}\\
  Heliophysics Sciences Division\\
  NASA Goddard Spaceflight Center\\
  Greenbelt, MD 20770 \\
  \texttt{michael.s.kirk@nasa.gov} \\
  \And
   Nat Mathews$^*$ \\
  Heliophysics Sciences Division\\
  NASA Goddard Spaceflight Center\\
  Greenbelt, MD 20770 \\
  \texttt{n.h.mathews@nasa.gov} \\
    \And
  Andrés Muñoz-Jaramillo $^*$\\
  Southwest Research Institute  \\
  Boulder, CO 80302\\
  \texttt{amunozj@boulder.swri.edu} 
}

\begin{document}

\maketitle

\begin{abstract}

 In this work, we introduce a novel generative denoising diffusion model for synthesizing the Sun’s three-dimensional coronal magnetic field, a complex and dynamic region characterized by evolving magnetic structures. Despite daily variability, these structures exhibit recurring patterns and long-term cyclic trends, presenting unique modeling challenges and opportunities at the intersection of physics and machine learning. Our generative approach employs an innovative architecture influenced by Spherical Fourier Neural Operators (SFNO), operating within the spherical harmonic domain, where the scalar field corresponds directly to the magnetic potential under physical constraints. We trained this model using an extensive dataset comprising 11.7 years of daily coupled simulations from the Air Force Data Assimilative Photospheric Flux Transport-Wang Sheeley Arge (ADAPT-WSA) model, further enhanced by data augmentation.  Initial results demonstrate the model's capability to conditionally generate physically realistic magnetic fields reflective of distinct phases within the 11-year solar cycle: from solar minimum ($S = 0$) to solar maximum ($S = 1$). This approach represents a significant step toward advanced generative three-dimensional modeling in Heliophysics, with potential applications in solar forecasting, data assimilation, inverse problem-solving, and broader impacts in areas such as procedural generation of physically-informed graphical assets.

  \end{abstract}

\section{Introduction}
The sun's corona is a region of intricate, time-dependent magnetic structures embedded in highly ionized plasmas. These three-dimensional magnetic configurations play a crucial role in shaping the solar wind, enabling the estimation of fundamental variables such as flow speed and mass density  \cite{arge2004stream,riley2015role,wallace2020relationship,gibson2016forward}. 

In this work, we detail a methodology for generative modeling of the three-dimensional magnetic field in the corona using denoising diffusion probabilistic modeling (DDPM) \cite{ho2020denoising,sohl2015deep}. Our method leverages 11.7 years of simulations spanning one complete solar cycle, generated using the Air Force Data Assimilative Photospheric Flux Transport-Wang Sheeley Arge (ADAPT-WSA) model.

Recent studies have demonstrated the effectiveness of DDPMs in modeling two-dimensional line-of-sight imagery of the solar corona  \cite{francisco2024generative,ramunno2024magnetogram}. This work builds upon these successes by expanding generative modeling to three-dimensional structures to advance solar physics research.

Source code developed for this project is publicly available at \url{https://github.com/ddasilva/coronal-diffusion-modeling/}. 

\subsection{Spherical Harmonics: Review for 3D Magnetism}

The Sun's corona can be modeled effectively using potential field simulations on time scales of a day or longer \cite{riley2006comparison}. These models utilize the assumption that $\nabla \times \mathbf{B} \approx 0$ within the near-sun environment, which gives rise to the scalar field $V$, representing magnetic potential, with the corresponding magnetic field modeled as $\mathbf{B}=-\nabla V$. 

To satisfy  Maxwell's law constraint  $\nabla \cdot \mathbf{B} = 0$, potential field simulations employ spherical harmonics as basis functions. These functions, combined with a radial component, satisfy Laplace's equation ($\nabla ^2 V = 0$), maintaining  $\nabla \cdot \mathbf{B} = 0$ \cite{whaler1981spherical}.

In this model, the three-dimensional magnetic field is represented using two lower-triangular matrices, $\mathbf{G}$ and $\mathbf{H}$, whose coefficients correspond to spherical harmonic basis functions.

\begin{figure}[h]
    \centering
    \includegraphics[width=1.0\textwidth]{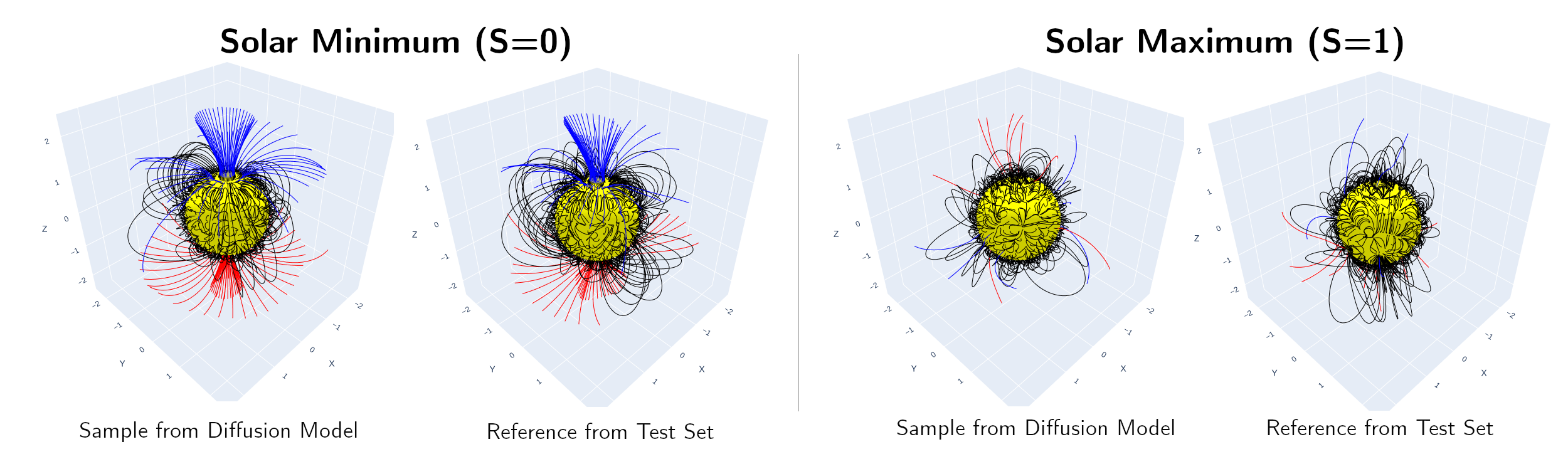}
    \caption{Samples from the model with references example from the test set. Solar minimum is called such because the field lines are more structured, symmetric, and  resemble a magnetic dipole. During solar maximum, the field lines are more chaotic and less like a dipole.
    \label{fig:fieldsamples}}
\end{figure}

\section{Data}

Simulation outputs between May 1, 2010, and December 31, 2021, at a daily cadence covering approximately one solar cycle are used. The simulations leverage ADAPT-WSA, with the ADAPT \cite{arge2010air,arge2011improving,arge2013modeling} model generating global surface magnetic field maps that serve as inputs to WSA \cite{sheeley2017origin}.  This data is part of the SuryaBench dataset \cite{roy2025suryabench}, publicly available under the MIT license at \url{https://huggingface.co/datasets/nasa-ibm-ai4science/surya-bench-coronal-extrapolation}.

ADAPT evolves the global magnetic field on the Sun's surface using flux transport models and continuously assimilates remote sensing measurements \cite{hickmann2015data,schrijver2003photospheric,worden2000evolving}. Remote sensing measurements assimilated are from the Solar Dynamics Observatory (SDO) Helioseismic and Magnetic Imager (HMI) surface vector magnetic field maps \cite{scherrer2012helioseismic}. All 12 realizations produced by the ADAPT ensemble are used for model training and testing. 

We use a subset of WSA that models the potential field between $1$ and $2.5~R_\odot$ (solar radii), where the magnetic field variability is most pronounced \cite{da2023ensemble}. WSA outputs include spherical harmonic coefficients $\mathbf{G}$ and $\mathbf{H}$ (degree 90), further augmented through vertical flips and $1^\circ$ azimuthal rotations \cite {wieczorek2018shtools}, ensuring Nyquist compliance. After augmentation, the training set consists of 27.3M items.

The solar cycle phase was quantified using the LISIRD Penticon F10.7 Solar Radio Flux Dataset \cite{leise2019lasp}. Values were processed through a $\pm 6$ month median filter to exclude outliers caused by solar flares. After filtering, the values were normalized to $[0, 1]$.

A common test/train split in solar physics was adopted, where January-August data was used for training and September-December data for validation.

\section{Methods}

\begin{figure}[h]
    \centering
    \includegraphics[width=0.5\textwidth]{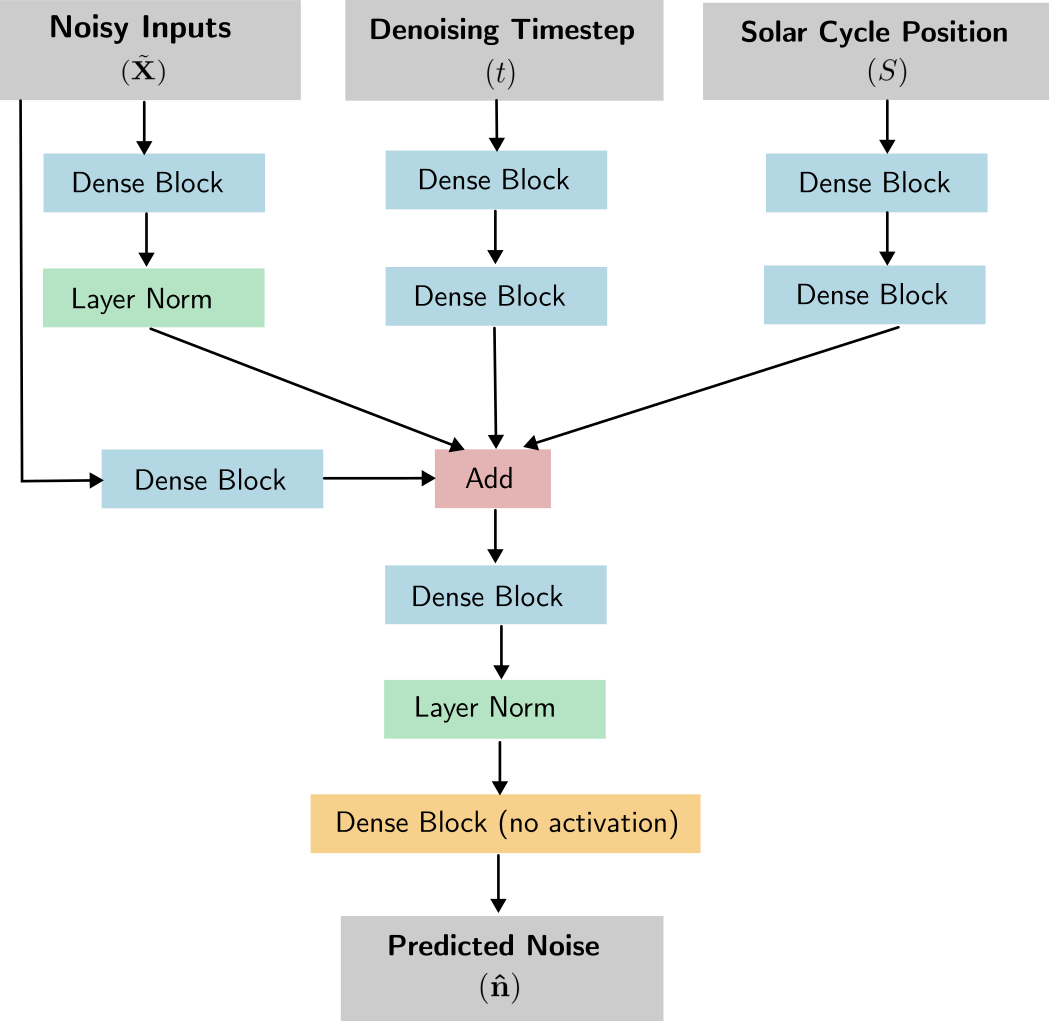}
    \caption{Neural Architecture, utilizing time embedding, context, and skip connections.
    \label{fig:netarch}}
\end{figure}

The DDPM model follows the scheme proposed by Ho et al., 2020 \cite{ho2020denoising}, where the network makes a prediction $\mathbf{\hat{n}}$ of the noise $\mathbf{n}$ mixed into the normalized authentic sample $\mathbf{X}$ at a rate assigned by a noise schedule. $\mathbf{X}$ is constructed by flattening the non-zero elements of $\mathbf{G}$ and $\mathbf{H}$. The noise is assigned according to a linear schedule parameterized by $\beta_1 = 10^{-4}$, $\beta_T=0.02$, and $T=1000$.


The network learns to map $(\tilde{\mathbf{X}}, t, S) \mapsto \hat{\mathbf{n}}$, where $\tilde{\mathbf{X}}$ is a perturbed sample,  $t\in[0,1]$ identifies noise schedule progress, and $S\in[0,1]$ reflects solar cycle position based on F10.7 flux data.

The network architecture includes time embedding, context embeddings ($S$), and skip connections (illustrated in Figure \ref{fig:netarch}). A dense block comprises a linear unit with an output dimension 8281 (size of $\mathbf{X}$) and a Leaky ReLU activation (negative slope of 0.1).

The loss function is the mean squared error within the spherical harmonic domain, which is the same domain in which the network operates.  This is inspired by Spherical Fourier Neural Operators (SNFO), an approach previously applied to auto-regressive time evolution on the sphere \cite{bonev2023spherical}. An advantage of using SNFO over e.g.  voxel vector fields, is that SFNO guarantees the results to be smooth.

Sampling employed DDIM \cite{song2020denoising} with $\eta = 0.1$ and $n=20$. We found that when starting sampling at $x_0 \sim \mathcal{N}(0, 1)$, the reverse diffusion trajectories became biased towards solar solar minimum irregardless of $S$. To solve this, seed helpers were created consisting of secondary means $\bar{x}_s$ and standard deviations $\sigma_s$ for subsets of the normalized test data in neighborhood of the intended trajectory ($S<0.25$ for solar minimum and $S>0.75$ for solar maximum). When sampling was done with using $x_0 \sim \mathcal{N}(\bar{x}_s, \sigma_s)$ from these seed helpers, trajectories more reliably followed the $S$ provided.

Training was performed using NVIDIA Quadro P5000 (16 GB of GPU memory) on a system with 32 GB of host memory using less than 8 CPUs. Augmenting required approximately 32 hours, while training took about \~48 hours to complete two epochs. A learning rate of .0001 was used with the learning rate decaying to 80\% of the initial value after the first epoch. Adam optimization \cite{kingma2014adam}. was used with a batch size of 128.  The total size of the weights on disk was 1.3 GB. 

\section{Results}

In Figure \ref{fig:fieldsamples}, we present two generated samples corresponding to solar minimum ($S=0$) and solar maximum ($S=1$). In these visualizations, red/blue lines denote open magnetic field lines (extending far into the solar system), while black lines denote closed field structures that trace backwards to the surface. 

\begin{figure}[h]
    \centering
    \includegraphics[width=0.8\textwidth]{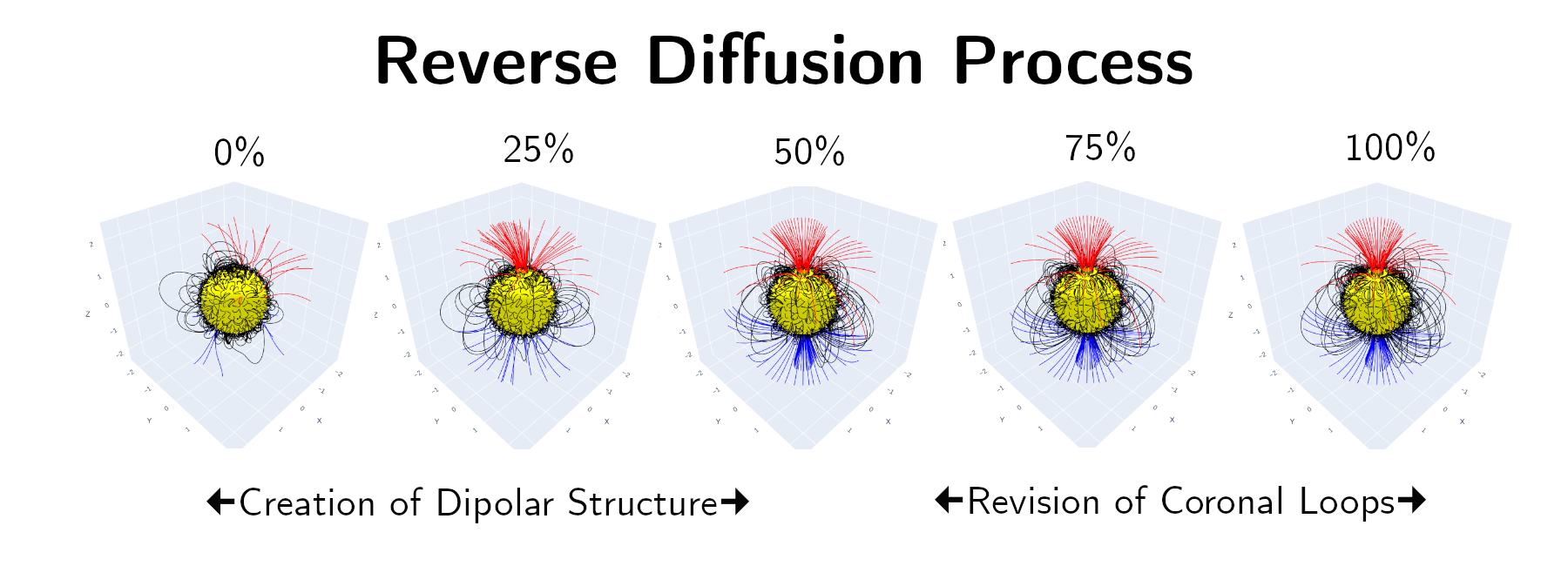}
    \caption{Illustration of the iterative reverse diffusion process for a $S=0$ sample.
    \label{fig:reverse_diffusion}}
\end{figure}

In Figure \ref{fig:reverse_diffusion}, we show strided steps of the DDIM reverse diffusion process performed to generate a sample with $S=0$. In the early stages of sample generation, the broad structure relating to a magnetic dipole is created and in later steps the structure of more localized coronal loops are refined. This is consistent with the patterns of generative diffusion models for photographic images \cite{ho2020denoising}, where-in early steps create broad shapes and later steps refine edges and textures.

\begin{figure}[h]
    \centering
    \includegraphics[width=\textwidth]{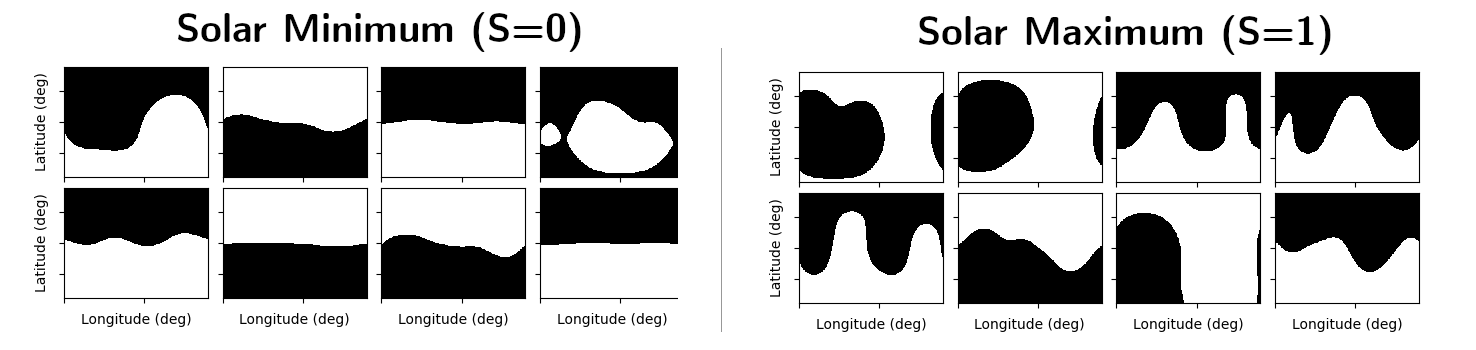}
    \caption{Plots of the Heliospheric Current Sheet (HCS) from random samples for $S=0$ and $S=1$.
    \label{fig:current}}
\end{figure}

Samples from the model we further evaluated in Figure \ref{fig:current} on the basis on a realisticness of a derived feature known as the heliospheric current sheet (HCS) \cite{smith2001heliospheric}. The HCS exists throughout the solar system, but at the outer edge of the corona (taken here at 5.0 solar radii), is the boundary which separates regions of positive (black) and negative (white) radial magnetic field components. Most of the generated samples gave rise to HCS configurations consistent with knowledge of the solar cycle. Specifically, in solar minimum the the HCS is mostly near the equator for solar minimum, and more dynamic and curved during solar maximum. 


This work is limited in that the only context is uses is the solar cycle position $S$. Future extensions to this work will include conditioning on additional parameters such as remote sensing observations (multispectral imagery of the corona and solar disk, and derived quantities like sunspot number).

\section{Conclusion and Outlook}

The field of space weather, broadly the effects of the sun and solar wind on technological systems, has a significant economic and societal impact. Space weather phenomena affect critical infrastructure, including power grids, communication networks, the GNSS systems, and aviation \cite{eastwood2017economic}.  

The results of this work highlight the potential for generative diffusion models to eventually become a powerful tool for solving classical problems in Heliophysics. For example, these models could provide unique responses to challenges in inverse modeling, data assimilation, and forecasting.  By learning physically informed distributions, diffusion models can aid in key advancements, such as improved solar wind predictions and magnetic field reconstructions. In space weather's sibling field of terrestrial weather forecasting,  diffusion models like GenCast \cite{price2025probabilistic} have demonstrated promising analog capability.  

Outside of pure research and forecasting, this work has applications in graphical asset generation for multimedia projects. For instance, video games or simulations could use this methodology to procedurally generate realistic sun-like objects informed by physical principles, enabling immersive and scientifically grounded environments within virtual universes.

\bibliographystyle{plain}
\bibliography{references}

\end{document}